\documentstyle{article}

\begin{document}

\title{THE LEPTON, QUARK AND HADRON CURRENTS}
\author{Gunn A.Quznetsov \\
quznets@geocities.com}
\maketitle

\begin{abstract}
The Clifford pentads of the 4X4 complex matrices define the current vectors
of the particles. The weak isospin transformation divides the particles on
two components, which scatter in the 2-dimensional antidiagonal Clifford
matrices space. A physics objects move in the 3-dimensional diagonal
Clifford matrices space. This sectioning of the 5-dimensional space on the
3-dimensional and the 2-dimensional subspaces defines the Newtonian gravity
principle.

The Clifford pentads sextet contains single light pentad and three chromatic
pentads. The Cartesian frame rotations confound the chromatic pentads. The
combination of the chromatic particles (the hadron monad) exists, which
behaves as the particle for such rotations .

\end{abstract}

\tableofcontents


\begin{quote}
Use the natural metric: $\hbar =c=1$.
\end{quote}

\section{ INTRODUCTION}

In the Quantum Theory the fermion behavior is depicted by the spinor $\Psi .$
The probability current vector $\overrightarrow{j}$ components of this
fermion are the following:

\begin{equation}
j_x=\Psi ^{\dagger }\cdot \beta ^1\cdot \Psi ,j_y=\Psi ^{\dagger }\cdot
\beta ^2\cdot \Psi ,j_z=\Psi ^{\dagger }\cdot \beta ^3\cdot \Psi .
\label{w0}
\end{equation}

Here

\[
\beta ^1=\left[ 
\begin{array}{cccc}
0 & 1 & 0 & 0 \\ 
1 & 0 & 0 & 0 \\ 
0 & 0 & 0 & -1 \\ 
0 & 0 & -1 & 0
\end{array}
\right] ,\beta ^2=\left[ 
\begin{array}{cccc}
0 & -i & 0 & 0 \\ 
i & 0 & 0 & 0 \\ 
0 & 0 & 0 & i \\ 
0 & 0 & -i & 0
\end{array}
\right] , 
\]

\[
\beta ^3=\left[ 
\begin{array}{cccc}
1 & 0 & 0 & 0 \\ 
0 & -1 & 0 & 0 \\ 
0 & 0 & -1 & 0 \\ 
0 & 0 & 0 & 1
\end{array}
\right] \mbox{.} 
\]

are the members of the Clifford pentad, for which other members are the
following:

\[
\gamma ^0=\left[ 
\begin{array}{cccc}
0 & 0 & 1 & 0 \\ 
0 & 0 & 0 & 1 \\ 
1 & 0 & 0 & 0 \\ 
0 & 1 & 0 & 0
\end{array}
\right] \mbox{ and }\beta ^4=\left[ 
\begin{array}{cccc}
0 & 0 & i & 0 \\ 
0 & 0 & 0 & i \\ 
-i & 0 & 0 & 0 \\ 
0 & -i & 0 & 0
\end{array}
\right] \mbox{.} 
\]

Let this spinor be expressed in the following form:

\[
\Psi =\left| \Psi \right| \cdot \left[ 
\begin{array}{c}
\exp \left( i\cdot g\right) \cdot \cos \left( b\right) \cdot \cos \left(
a\right) \\ 
\exp \left( i\cdot d\right) \cdot \sin \left( b\right) \cdot \cos \left(
a\right) \\ 
\exp \left( i\cdot f\right) \cdot \cos \left( v\right) \cdot \sin \left(
a\right) \\ 
\exp \left( i\cdot q\right) \cdot \sin \left( v\right) \cdot \sin \left(
a\right)
\end{array}
\right] . 
\]

In this case the probability current vector $\overrightarrow{j}$ has got the
following components:

\begin{equation}
\begin{array}{c}
j_x=\left| \Psi \right| ^2\cdot \\ 
\cdot \left[ \cos ^2\left( a\right) \cdot \sin \left( 2\cdot b\right) \cdot
\cos \left( d-g\right) -\sin ^2\left( a\right) \cdot \sin \left( 2\cdot
v\right) \cdot \cos \left( q-f\right) \right] , \\ 
j_y=\left| \Psi \right| ^2\cdot \\ 
\cdot \left[ \cos ^2\left( a\right) \cdot \sin \left( 2\cdot b\right) \cdot
\sin \left( d-g\right) -\sin ^2\left( a\right) \cdot \sin \left( 2\cdot
v\right) \cdot \sin \left( q-f\right) \right] , \\ 
j_z=\left| \Psi \right| ^2\cdot \left[ \cos ^2\left( a\right) \cdot \cos
\left( 2\cdot b\right) -\sin ^2\left( a\right) \cdot \cos \left( 2\cdot
v\right) \right] .
\end{array}
\label{w1}
\end{equation}

If

\[
\rho =\Psi ^{\dagger }\cdot \Psi \mbox{, } 
\]

then $\rho $ is the probability density, i.e.$\int \int \int_{\left(
V\right) }\rho \left( t\right) \cdot dV$ is the probability to find the
particle with the state function $\Psi $ in the domain $V$ of the
3-dimensional space at the time moment $t$. In this case, $\{\rho ,%
\overrightarrow{j}\}$ is the probability density $3+1$-vector.

If

\begin{equation}
\overrightarrow{j}=\rho \cdot \overrightarrow{u},  \label{w2}
\end{equation}

then $\overrightarrow{u}$ is the average velocity for this particle.

Let us denote:

\begin{equation}
J_0=\Psi ^{\dagger }\cdot \gamma ^0\cdot \Psi ,J_4=\Psi ^{\dagger }\cdot
\beta ^4\cdot \Psi ,J_0=\rho \cdot V_0,J_4=\rho \cdot V_4.  \label{w3}
\end{equation}

In this case:

\begin{equation}
\begin{array}{c}
V_0=\sin \left( 2\cdot a\right) \cdot \left[ \cos \left( b\right) \cdot \cos
\left( v\right) \cdot \cos \left( g-f\right) +\sin \left( b\right) \cdot
\sin \left( v\right) \cdot \cos \left( d-q\right) \right] \mbox{,} \\ 
V_4=\sin \left( 2\cdot a\right) \cdot \left[ \cos \left( b\right) \cdot \cos
\left( v\right) \cdot \sin \left( g-f\right) +\sin \left( b\right) \cdot
\sin \left( v\right) \cdot \sin \left( d-q\right) \right] ;
\end{array}
\label{w4}
\end{equation}

and for every particle:

\begin{equation}
u_x^2+u_y^2+u_z^2+V_0^2+V_4^2=1.  \label{VV}
\end{equation}

For the left particle (for example, the left neutrino): $a=\frac \pi 2$,

\[
\Psi _L=\left| \Psi _L\right| \cdot \left[ 
\begin{array}{c}
0 \\ 
0 \\ 
\exp \left( i\cdot f\right) \cdot \cos \left( v\right) \\ 
\exp \left( i\cdot q\right) \cdot \sin \left( v\right)
\end{array}
\right] 
\]

and from ($\ref{w1}$), and ($\ref{w2}$): $u_x^2+u_y^2+u_z^2=1$.Hence, the
left particle velocity equals $1$; hence, the mass of the left particle
equals to zero.

Let $U$ be the weak global isospin (SU(2)) transformation with the
eigenvalues $\exp \left( i\cdot \lambda \right) $.

In this case for this transformation eigenvector $\psi $:

\[
U\psi =\left| \psi \right| \cdot \left[ 
\begin{array}{c}
\exp \left( i\cdot g\right) \cdot \cos \left( b\right) \cdot \cos \left(
a\right) \\ 
\exp \left( i\cdot d\right) \cdot \sin \left( b\right) \cdot \cos \left(
a\right) \\ 
\exp \left( i\cdot \lambda \right) \cdot \exp \left( i\cdot f\right) \cdot
\cos \left( v\right) \cdot \sin \left( a\right) \\ 
\exp \left( i\cdot \lambda \right) \cdot \exp \left( i\cdot q\right) \cdot
\sin \left( v\right) \cdot \sin \left( a\right)
\end{array}
\right] 
\]

and for $1\leq \mu \leq 3$ from (\ref{w1}):

\begin{equation}
\left( U\psi \right) ^{\dagger }\cdot \beta ^\mu \cdot \left( U\psi \right)
=\psi ^{\dagger }\cdot \beta ^\mu \cdot \psi ,  \label{w5}
\end{equation}

but for $\mu =0$ and $\mu =4$ from (\ref{w4}):

\begin{equation}
\begin{array}{c}
\begin{array}{c}
\psi ^{\dagger }\cdot \gamma ^0\cdot \psi =\left| \psi \right| ^2\cdot \sin
\left( 2\cdot a\right) \cdot \\ 
\left[ \cos \left( b\right) \cdot \cos \left( v\right) \cdot \cos \left(
g-f-\lambda \right) +\sin \left( b\right) \cdot \sin \left( v\right) \cdot
\cos \left( d-q-\lambda \right) \right] \mbox{, }
\end{array}
\\ 
\begin{array}{c}
\psi ^{\dagger }\cdot \beta ^4\cdot \psi =\left| \psi \right| ^2\cdot \sin
\left( 2\cdot a\right) \cdot \\ 
\left[ \cos \left( b\right) \cdot \cos \left( v\right) \cdot \sin \left(
g-f-\lambda \right) +\sin \left( b\right) \cdot \sin \left( v\right) \cdot
\sin \left( d-q-\lambda \right) \right] \mbox{; }
\end{array}
\end{array}
\label{w6}
\end{equation}

\section{ THE WEAK ISOSPIN SPACE}

In the weak isospin theory we have got the following entities ({Global
Symmetries, Standard Model}):

the right electron state vector $e_R$,

the left electron state vector $e_L$,

the electron state vector $e$ ($e=\left[ 
\begin{array}{c}
e_R \\ 
e_L
\end{array}
\right] $),

the left neutrino state vector $\nu _L$,

the zero vector right neutrino $\nu _R$.

the unitary $2\times 2$ matrix $U$ of the isospin transformation.($\det
\left( U\right) =1$) ({Gauge Symmetry}).

This matrix acts on the vectors of the kind:$\left[ 
\begin{array}{c}
\nu _L \\ 
e_L
\end{array}
\right] $.

Therefore, in this theory: if

\[
U=\left[ 
\begin{array}{cc}
u_{1,1} & u_{1,2} \\ 
u_{2,1} & u_{2,2}
\end{array}
\right] 
\]

then the matrix

\[
\left[ 
\begin{array}{cccc}
1 & 0 & 0 & 0 \\ 
0 & u_{1,1} & 0 & u_{1,2} \\ 
0 & 0 & 1 & 0 \\ 
0 & u_{2,1} & 0 & u_{2,2}
\end{array}
\right] 
\]

operates on the vector

\[
\left[ 
\begin{array}{c}
e_R \\ 
e_L \\ 
\nu _R \\ 
\nu _L
\end{array}
\right] \mbox{.} 
\]

Because $e_R$, $e_L$, $\nu _R$, $\nu _L$ are the two-component vectors then

\[
\left[ 
\begin{array}{c}
e_R \\ 
e_L \\ 
\nu _R \\ 
\nu _L
\end{array}
\right] \mbox{ is }\left[ 
\begin{array}{c}
e_{R1} \\ 
e_{R2} \\ 
e_{L1} \\ 
e_{L2} \\ 
0 \\ 
0 \\ 
\nu _{L1} \\ 
\nu _{L2}
\end{array}
\right] 
\]

and

\[
\left[ 
\begin{array}{cccc}
1 & 0 & 0 & 0 \\ 
0 & u_{1,1} & 0 & u_{1,2} \\ 
0 & 0 & 1 & 0 \\ 
0 & u_{2,1} & 0 & u_{2,2}
\end{array}
\right] \mbox{ is }\underline{U}=\mbox{ }\left[ 
\begin{array}{cccccccc}
1 & 0 & 0 & 0 & 0 & 0 & 0 & 0 \\ 
0 & 1 & 0 & 0 & 0 & 0 & 0 & 0 \\ 
0 & 0 & u_{1,1} & 0 & 0 & 0 & u_{1,2} & 0 \\ 
0 & 0 & 0 & u_{1,1} & 0 & 0 & 0 & u_{1,2} \\ 
0 & 0 & 0 & 0 & 1 & 0 & 0 & 0 \\ 
0 & 0 & 0 & 0 & 0 & 1 & 0 & 0 \\ 
0 & 0 & u_{2,1} & 0 & 0 & 0 & u_{2,2} & 0 \\ 
0 & 0 & 0 & u_{2,1} & 0 & 0 & 0 & u_{2,2}
\end{array}
\right] \mbox{.} 
\]

This matrix has got eight orthogonal normalized eigenvectors of kind:

\[
\underline{s_1}=\left[ 
\begin{array}{c}
1 \\ 
0 \\ 
0 \\ 
0 \\ 
0 \\ 
0 \\ 
0 \\ 
0
\end{array}
\right] ,\underline{s_2}=\left[ 
\begin{array}{c}
0 \\ 
1 \\ 
0 \\ 
0 \\ 
0 \\ 
0 \\ 
0 \\ 
0
\end{array}
\right] ,\underline{s_3}=\left[ 
\begin{array}{c}
0 \\ 
0 \\ 
\varpi \\ 
0 \\ 
0 \\ 
0 \\ 
\chi \\ 
0
\end{array}
\right] ,\underline{s_4}=\left[ 
\begin{array}{c}
0 \\ 
0 \\ 
0 \\ 
\chi ^{*} \\ 
0 \\ 
0 \\ 
0 \\ 
-\varpi ^{*}
\end{array}
\right] , 
\]

\[
\underline{s_5}=\left[ 
\begin{array}{c}
0 \\ 
0 \\ 
0 \\ 
0 \\ 
1 \\ 
0 \\ 
0 \\ 
0
\end{array}
\right] ,\underline{s_6}=\left[ 
\begin{array}{c}
0 \\ 
0 \\ 
0 \\ 
0 \\ 
0 \\ 
1 \\ 
0 \\ 
0
\end{array}
\right] ,\underline{s_7}=\left[ 
\begin{array}{c}
0 \\ 
0 \\ 
\chi ^{*} \\ 
0 \\ 
0 \\ 
0 \\ 
-\varpi ^{*} \\ 
0
\end{array}
\right] ,\underline{s_8}=\left[ 
\begin{array}{c}
0 \\ 
0 \\ 
0 \\ 
\varpi \\ 
0 \\ 
0 \\ 
0 \\ 
\chi
\end{array}
\right] . 
\]

The corresponding eigenvalues are: $1$, $1$, $\exp \left( i\cdot \lambda
\right) $, $\exp \left( i\cdot \lambda \right) $, $1$,$1$,

$\exp \left( -i\cdot \lambda \right) $, $\exp \left( -i\cdot \lambda \right) 
$.

These vectors constitute the orthogonal basis in this 8-dimensional space.

Let $\underline{\gamma ^0}=\left[ 
\begin{array}{cc}
\gamma ^0 & O \\ 
O & \gamma ^0
\end{array}
\right] $, if $O$ is zero $4\times 4$ matrix, and \underline{$\beta ^4$}$%
=\left[ 
\begin{array}{cc}
\beta ^4 & O \\ 
O & \beta ^4
\end{array}
\right] $.

The vectors $\left[ 
\begin{array}{c}
e_{R1} \\ 
e_{R2} \\ 
e_{L1} \\ 
e_{L2} \\ 
0 \\ 
0 \\ 
0 \\ 
0
\end{array}
\right] $, $\left[ 
\begin{array}{c}
e_{R1} \\ 
e_{R2} \\ 
0 \\ 
0 \\ 
0 \\ 
0 \\ 
0 \\ 
0
\end{array}
\right] $, $\left[ 
\begin{array}{c}
0 \\ 
0 \\ 
e_{L1} \\ 
e_{L2} \\ 
0 \\ 
0 \\ 
0 \\ 
0
\end{array}
\right] $ correspond to the state vectors $e$, $e_R$ and $e_L$ resp.

In this case (\ref{w3}) \underline{$e$}$^{\dagger }\cdot \underline{\gamma ^0%
}\cdot \underline{e}=J_{0e}$, \underline{$e$}$^{\dagger }\cdot \underline{%
\beta ^4}\cdot \underline{e}=J_{4e}$, $J_{0e}=\underline{e}^{\dagger }\cdot 
\underline{e}\cdot V_{0e}$, $J_{4e}=\underline{e}^{\dagger }\cdot \underline{%
e}\cdot V_{4e}$.

For the vector \underline{$e$} the numbers $k_3$, $k_4$, $k_7$, $k_8$ exist,
for which: \underline{$e$}$=(e_{R1}\cdot \underline{s_1}+e_{R2}\cdot 
\underline{s_2})+(k_3\cdot \underline{s_3}+k_4\cdot \underline{s_4}%
)+(k_7\cdot \underline{s_7}+k_8\cdot \underline{s_8})$.

Here \underline{$e_R$}$=(e_{R1}\cdot \underline{s_1}+e_{R2}\cdot \underline{%
s_2})$. If \underline{$e_{La}$}$=(k_3\cdot \underline{s_3}+k_4\cdot 
\underline{s_4})$ and \underline{$e_{Lb}$}$=(k_7\cdot \underline{s_7}%
+k_8\cdot \underline{s_8})$ then \underline{$U$}$\cdot \underline{e_{La}}%
=\exp \left( i\cdot \lambda \right) \cdot $\underline{$e_{La}$} and 
\underline{$U$}$\cdot \underline{e_{Lb}}=\exp \left( -i\cdot \lambda \right)
\cdot $\underline{$e_{Lb}$}.

Let for all $k$ ($1\leq k\leq 8$): \underline{$h_k$}$=\underline{\gamma ^0}%
\cdot $\underline{$s_k$}. The vectors \underline{$h_k$} constitute the
orthogonal basis, too. And the numbers $q_3$, $q_4$, $q_7$, $q_8$ exist, for
which: \underline{$e_R$}$=(q_3\cdot \underline{h_3}+q_4\cdot \underline{h_4}%
)+(q_7\cdot \underline{h_7}+q_8\cdot \underline{h_8})$.

Let \underline{$e_{Ra}$}$=(q_3\cdot \underline{h_3}+q_4\cdot \underline{h_4}%
) $, \underline{$e_{Rb}$}$=(q_7\cdot \underline{h_7}+q_8\cdot \underline{h_8}%
)$, \underline{$e_a$}$=\underline{e_{Ra}}+$\underline{$e_{La}$} and 
\underline{$e_b$}$=\underline{e_{Rb}}+$\underline{$e_{Lb}$}.

Let \underline{$e_a$}$^{\dagger }\cdot \underline{\gamma ^0}\cdot \underline{%
e_a}=J_{0a}$, \underline{$e_a$}$^{\dagger }\cdot \underline{\beta ^4}\cdot 
\underline{e_a}=J_{4a}$, $J_{0a}=\underline{e_a}^{\dagger }\cdot \underline{%
e_a}\cdot V_{0a}$, $J_{4a}=\underline{e_a}^{\dagger }\cdot \underline{e_a}%
\cdot V_{4a}$,

\underline{$e_b$}$^{\dagger }\cdot \underline{\gamma ^0}\cdot \underline{e_b}%
=J_{0b}$, \underline{$e_b$}$^{\dagger }\cdot \underline{\beta ^4}\cdot 
\underline{e_b}=J_{4b}$, $J_{0b}=\underline{e_b}^{\dagger }\cdot \underline{%
e_b}\cdot V_{0b}$, $J_{4b}=\underline{e_b}^{\dagger }\cdot \underline{e_b}%
\cdot V_{4b}$.

In this case: $J_0=J_{0a}+J_{0b}$, $J_4=J_{4a}+J_{4b}$.

Let $\left( \underline{U}\cdot \underline{e_a}\right) ^{\dagger }\cdot 
\underline{\gamma ^0}\cdot \left( \underline{U}\cdot \underline{e_a}\right)
=J_{0a}^{\prime }$, $\left( \underline{U}\cdot \underline{e_a}\right)
^{\dagger }\cdot \underline{\beta ^4}\cdot \underline{\left( \underline{U}%
\cdot \underline{e_a}\right) }=J_{4a}^{\prime }$, $J_{0a}^{\prime }=\left( 
\underline{U}\cdot \underline{e_a}\right) ^{\dagger }\cdot \left( \underline{%
U}\cdot \underline{e_a}\right) \cdot V_{0a}^{\prime }$, $J_{4a}^{\prime
}=\left( \underline{U}\cdot \underline{e_a}\right) ^{\dagger }\cdot \left( 
\underline{U}\cdot \underline{e_a}\right) \cdot V_{4a}^{\prime }$,

$\left( \underline{U}\cdot \underline{e_b}\right) ^{\dagger }\cdot 
\underline{\gamma ^0}\cdot \left( \underline{U}\cdot \underline{e_b}\right)
=J_{0b}^{\prime }$, $\left( \underline{U}\cdot \underline{e_b}\right)
^{\dagger }\cdot \underline{\beta ^4}\cdot \left( \underline{U}\cdot 
\underline{e_b}\right) =J_{4b}^{\prime }$, $J_{0b}^{\prime }=\left( 
\underline{U}\cdot \underline{e_b}\right) ^{\dagger }\cdot \left( \underline{%
U}\cdot \underline{e_b}\right) \cdot V_{0b}^{\prime }$, $J_{4b}^{\prime
}=\left( \underline{U}\cdot \underline{e_b}\right) ^{\dagger }\cdot \left( 
\underline{U}\cdot \underline{e_b}\right) \cdot V_{4b}^{\prime }$.

In this case from (\ref{w6}):

\[
\begin{array}{c}
V_{0a}^{\prime }=V_{0a}\cdot \cos \left( \lambda \right) +V_{4a}\cdot \sin
\left( \lambda \right) , \\ 
V_{4a}^{\prime }=V_{4a}\cdot \cos \left( \lambda \right) -V_{0a}\cdot \sin
\left( \lambda \right) ; \\ 
V_{0b}^{\prime }=V_{0b}\cdot \cos \left( \lambda \right) -V_{4b}\cdot \sin
\left( \lambda \right) , \\ 
V_{4b}^{\prime }=V_{4b}\cdot \cos \left( \lambda \right) +V_{0b}\cdot \sin
\left( \lambda \right) \mbox{.}
\end{array}
\]

Hence, every isospin transformation divides a electron on two components,
which scatter on the angle $2\cdot \lambda $ in the space of ($J_0$, $J_4$).

These components are indiscernible in the space of ($j_x$, $j_y$, $j_z$) (%
\ref{w5}).

Let $o$ be the $2\times 2$ zeros matrix. Let the $4\times 4$ matrices of
kind:

\[
\left[ 
\begin{array}{cc}
P & o \\ 
o & S
\end{array}
\right] 
\]

be denoted as the diagonal matrices, and

\[
\left[ 
\begin{array}{cc}
o & P \\ 
S & o
\end{array}
\right] 
\]

be denoted as the antidiogonal matrices.

Three diagonal members ($\beta ^1$, $\beta ^2$, $\beta ^3$) of the Clifford
pentad define the 3-dimensional space $\Re $, in which $u_x$, $u_y$ , $u_z$
are located. The physics objects move in this space. Two antidiogonal
members ($\gamma ^0$, $\beta ^4$) of this pentad define the 2-dimensional
space $\grave A$, in which $V_0$ and $V_4$ are located. The weak isospin
transformation acts in this space.

\section{GRAVITY}

Let $x$ be the particle average coordinate in $\Re $, and ${\bf m}$ be the
average coordinate of this particle in $\grave A$. Let $x+i\cdot {\bf m}$ be
denoted as the complex coordinate of this particle.

From (\ref{VV}) this particle average velocity, which proportional to $%
x+i\cdot {\bf m}$, is:

\[
v=\frac{x+i\cdot {\bf m}}{\sqrt{\left( x^2+{\bf m}^2\right) }}\mbox{.} 
\]

$\left| v\right| =1$, but for the acceleration:

\[
a=\frac{dv}{dt}=\frac{dv}{dx}\cdot v=-i\cdot {\bf m\cdot }\left( \frac{%
x+i\cdot {\bf m}}{x^2+{\bf m}^2}\right) ^2\mbox{.} 
\]

And if ${\bf m}\ll x$, then

\[
\left| a\right| \simeq \frac{{\bf m}}{x^2}\mbox{.} 
\]

This is very much reminds the Newtonian gravity principle (Classical
Theories of Gravity).

\section{ THE CHROMATIC SPACE}

Let

\[
I=\left[ 
\begin{array}{cc}
1 & 0 \\ 
0 & 1
\end{array}
\right] 
\]

and

\[
\sigma _x=\left( 
\begin{array}{cc}
0 & 1 \\ 
1 & 0
\end{array}
\right) ,\sigma _y=\left( 
\begin{array}{cc}
0 & -i \\ 
i & 0
\end{array}
\right) ,\sigma _z=\left( 
\begin{array}{cc}
1 & 0 \\ 
0 & -1
\end{array}
\right) 
\]

be the Pauli matrices.

Six Clifford's pentads exists, only:

the red pentad $\zeta $:

\[
\zeta ^x=\left[ 
\begin{array}{cc}
\sigma _x & o \\ 
o & -\sigma _x
\end{array}
\right] ,\zeta ^y=\left[ 
\begin{array}{cc}
\sigma _y & o \\ 
o & \sigma _y
\end{array}
\right] ,\zeta ^z=\left[ 
\begin{array}{cc}
-\sigma _z & o \\ 
o & -\sigma _z
\end{array}
\right] , 
\]

\[
\gamma _\zeta ^0=\left[ 
\begin{array}{cc}
o & -\sigma _x \\ 
-\sigma _x & o
\end{array}
\right] \mbox{, }\zeta ^4=-i\cdot \left[ 
\begin{array}{cc}
o & \sigma _x \\ 
-\sigma _x & o
\end{array}
\right] ; 
\]

the green pentad $\eta $:

\[
\eta ^x=\left[ 
\begin{array}{cc}
-\sigma _x & o \\ 
o & -\sigma _x
\end{array}
\right] ,\eta ^y=\left[ 
\begin{array}{cc}
\sigma _y & o \\ 
o & -\sigma _y
\end{array}
\right] ,\eta ^z=\left[ 
\begin{array}{cc}
-\sigma _z & o \\ 
o & -\sigma _z
\end{array}
\right] , 
\]

\[
\gamma _\eta ^0=\left[ 
\begin{array}{cc}
o & -\sigma _y \\ 
-\sigma _y & o
\end{array}
\right] \mbox{, }\eta ^4=i\cdot \left[ 
\begin{array}{cc}
o & \sigma _y \\ 
-\sigma _y & o
\end{array}
\right] ; 
\]

the blue pentad $\theta $:

\[
\theta ^x=\left[ 
\begin{array}{cc}
-\sigma _x & o \\ 
o & -\sigma _x
\end{array}
\right] ,\theta ^y=\left[ 
\begin{array}{cc}
\sigma _y & o \\ 
o & \sigma _y
\end{array}
\right] ,\theta ^z=\left[ 
\begin{array}{cc}
\sigma _z & o \\ 
o & -\sigma _z
\end{array}
\right] , 
\]

\[
\gamma _\theta ^0=\left[ 
\begin{array}{cc}
o & -\sigma _z \\ 
-\sigma _z & o
\end{array}
\right] ,\theta ^4=-i\cdot \left[ 
\begin{array}{cc}
o & \sigma _z \\ 
-\sigma _z & o
\end{array}
\right] ; 
\]

the light pentad $\beta $:

\[
\beta ^x=\left[ 
\begin{array}{cc}
\sigma _x & o \\ 
o & -\sigma _x
\end{array}
\right] ,\beta ^y=\left[ 
\begin{array}{cc}
\sigma _y & o \\ 
o & -\sigma _y
\end{array}
\right] ,\beta ^z=\left[ 
\begin{array}{cc}
\sigma _z & o \\ 
o & -\sigma _z
\end{array}
\right] , 
\]

\[
\gamma ^0=\left[ 
\begin{array}{cc}
o & I \\ 
I & o
\end{array}
\right] ,\beta ^4=i\cdot \left[ 
\begin{array}{cc}
o & I \\ 
-I & o
\end{array}
\right] ; 
\]

the sweet pentad \underline{$\Delta $}:

\[
\underline{\Delta }^1=\left[ 
\begin{array}{cc}
o & -\sigma _x \\ 
-\sigma _x & o
\end{array}
\right] ,\underline{\Delta }^2=\left[ 
\begin{array}{cc}
o & -\sigma _y \\ 
-\sigma _y & o
\end{array}
\right] ,\underline{\Delta }^3=\left[ 
\begin{array}{cc}
o & -\sigma _z \\ 
-\sigma _z & o
\end{array}
\right] , 
\]

\[
\underline{\Delta }^0=\left[ 
\begin{array}{cc}
-I & o \\ 
o & I
\end{array}
\right] ,\underline{\Delta }^4=i\cdot \left[ 
\begin{array}{cc}
o & I \\ 
-I & o
\end{array}
\right] ; 
\]

the bitter pentad \underline{$\Gamma $}:

\[
\underline{\Gamma }^1=i\cdot \left[ 
\begin{array}{cc}
o & -\sigma _x \\ 
\sigma _x & o
\end{array}
\right] ,\underline{\Gamma }^2=i\cdot \left[ 
\begin{array}{cc}
o & -\sigma _y \\ 
\sigma _y & o
\end{array}
\right] ,\underline{\Gamma }^3=i\cdot \left[ 
\begin{array}{cc}
o & -\sigma _z \\ 
\sigma _z & o
\end{array}
\right] , 
\]

\[
\underline{\Gamma }^0=\left[ 
\begin{array}{cc}
-I & o \\ 
o & I
\end{array}
\right] ,\underline{\Gamma }^4=\left[ 
\begin{array}{cc}
o & I \\ 
I & o
\end{array}
\right] \mbox{.} 
\]

The average velocity vector for the sweet pentad has gotten the following
components:

\[
V_0^{\underline{\Delta }}=-\cos \left( 2\cdot a\right) , 
\]

\[
V_1^{\underline{\Delta }}=-\sin \left( 2\cdot a\right) \cdot \left[ \cos
\left( b\right) \cdot \sin \left( v\right) \cos \left( g-q\right) +\sin
\left( b\right) \cdot \cos \left( v\right) \cos \left( d-f\right) \right] , 
\]

\[
V_2^{\underline{\Delta }}=-\sin \left( 2\cdot a\right) \cdot \left[ -\cos
\left( b\right) \cdot \sin \left( v\right) \sin \left( g-q\right) +\sin
\left( b\right) \cdot \cos \left( v\right) \sin \left( d-f\right) \right] , 
\]

\[
V_3^{\underline{\Delta }}=-\sin \left( 2\cdot a\right) \cdot \left[ \cos
\left( b\right) \cdot \cos \left( v\right) \cos \left( g-f\right) -\sin
\left( b\right) \cdot \sin \left( v\right) \cos \left( d-q\right) \right] , 
\]

\[
V_4^{\underline{\Delta }}=-\sin \left( 2\cdot a\right) \cdot \left[ -\cos
\left( b\right) \cdot \cos \left( v\right) \sin \left( g-f\right) -\sin
\left( b\right) \cdot \sin \left( v\right) \sin \left( d-q\right) \right] . 
\]

Therefore, here the antidiogonal matrices $\underline{\Delta }^1$ and $%
\underline{\Delta }^2$ define the 2- dimensional space ($V_1^{\underline{%
\Delta }}$, $V_2^{\underline{\Delta }}$) in which the weak isospin
transformation acts. The antidiogonal matrices $\underline{\Delta }^3$ and $%
\underline{\Delta }^4$ define similar space ($V_3^{\underline{\Delta }}$, $%
V_4^{\underline{\Delta }}$). The sweet pentad is kept a single diagonal
matrix, which defines the one-dimensional space ($V_0^{\underline{\Delta }}$%
) for the moving of the objects.

Like the sweet pentad, the bitter pentad with four antidiogonal matrices and
with single diagonal matrix defines two 2-dimensional spaces, in which the
weak isospin transformation acts, and single one-dimensional space for the
moving of the objects.

Each chromatic pentad with 3 diagonal matrices and with 2 antidiogonal
matrices, like the light pentad, defines single 2-dimensional space, in
which the weak isospin interaction acts, and defines single 3-dimensional
space for the moving the physics objects.

Let $\phi _3$ be any real number and:

\[
\begin{array}{c}
x^{\prime }=x\cdot \cos \left( 2\cdot \phi _3\right) -y\cdot \sin \left(
2\cdot \phi _3\right) \mbox{,} \\ 
y^{\prime }=y\cdot \cos \left( 2\cdot \phi _3\right) +x\cdot \sin \left(
2\cdot \phi _3\right) \mbox{,} \\ 
z^{\prime }=z\mbox{.}
\end{array}
\]

That is the Cartesian frame $\left\{ x^{\prime },y^{\prime },z^{\prime
}\right\} $ is obtained from $\left\{ x,y,z\right\} $ by the rotation about
Z-axis on the angle $2\cdot \phi _3$.

In this case the velocity coordinates are transformed as the following:

\begin{equation}
\begin{array}{c}
u_x^{\prime }=u_x\cdot \cos \left( 2\cdot \phi _3\right) +u_y\cdot \sin
\left( 2\cdot \phi _3\right) \mbox{,} \\ 
u_y^{\prime }=u_y\cdot \cos \left( 2\cdot \phi _3\right) -u_x\cdot \sin
\left( 2\cdot \phi _3\right) \mbox{,} \\ 
u_z^{\prime }=u_z
\end{array}
\label{w7}
\end{equation}

If 
\[
U_z=-i\cdot \beta ^x\cdot \beta ^y\mbox{, }Q_z\left( \phi _3\right) =\cos
\left( \phi _3\right) \cdot E+i\cdot \sin \left( \phi _3\right) \cdot U_z%
\mbox{,} 
\]

and

\[
\Psi ^{\prime }=Q_z\left( \phi _3\right) \cdot \Psi \mbox{,} 
\]

then the light particles velocity fulfils to (\ref{w7}). And for these
particles: $V_0^{\beta \prime }=V_0^\beta $, $V_4^{\beta \prime }=V_4^\beta $%
.

Hence, $Q_z\left( \phi _3\right) $ coordinates to the Cartesian frame
rotation about Z-axis on the angle $2\cdot \phi _3$, because, if for all $%
\vartheta $: $\vartheta ^{\prime }=Q_z\left( \phi _3\right) ^{\dagger }\cdot
\vartheta \cdot Q_z\left( \phi _3\right) $, then

\[
\begin{array}{c}
\beta ^{x\prime }=\beta ^x\cdot \cos \left( 2\cdot \phi _3\right) -\beta
^y\cdot \sin \left( 2\cdot \phi _3\right) \mbox{,} \\ 
\beta ^{y\prime }=\beta ^y\cdot \cos \left( 2\cdot \phi _3\right) +\beta
^x\cdot \sin \left( 2\cdot \phi _3\right) \mbox{,} \\ 
\beta ^{z\prime }=\beta ^z\mbox{, } \\ 
\gamma ^{0\prime }=\gamma ^0\mbox{,} \\ 
\beta ^{4\prime }=\beta ^4\mbox{.}
\end{array}
\]

(\ref{w0}).

But

\[
\begin{array}{c}
\zeta ^{x\prime }=\zeta ^x\cdot \cos \left( 2\cdot \phi _3\right) +\eta
^y\cdot \sin \left( 2\cdot \phi _3\right) \mbox{,} \\ 
\eta ^{y\prime }=\eta ^y\cdot \cos \left( 2\cdot \phi _3\right) -\zeta
^x\cdot \sin \left( 2\cdot \phi _3\right) \mbox{,} \\ 
\zeta ^{y\prime }=\zeta ^y\cdot \cos \left( 2\cdot \phi _3\right) +\eta
^x\cdot \sin \left( 2\cdot \phi _3\right) \mbox{,} \\ 
\eta ^{x\prime }=\eta ^x\cdot \cos \left( 2\cdot \phi _3\right) -\zeta
^y\cdot \sin \left( 2\cdot \phi _3\right) \mbox{,} \\ 
\gamma _\zeta ^{0\prime }=\gamma _\zeta ^0\cdot \cos \left( 2\cdot \phi
_3\right) +\gamma _\eta ^0\cdot \sin \left( 2\cdot \phi _3\right) \mbox{,}
\\ 
\gamma _\eta ^{0\prime }=\gamma _\eta ^0\cdot \cos \left( 2\cdot \phi
_3\right) -\gamma _\zeta ^0\cdot \sin \left( 2\cdot \phi _3\right) \mbox{,}
\\ 
\zeta ^{4\prime }=\zeta ^4\cdot \cos \left( 2\cdot \phi _3\right) -\eta
^4\cdot \sin \left( 2\cdot \phi _3\right) \mbox{,} \\ 
\eta ^{4\prime }=\eta ^4\cdot \cos \left( 2\cdot \phi _3\right) +\zeta
^4\cdot \sin \left( 2\cdot \phi _3\right) \mbox{.}
\end{array}
\]

That is the red pentad and the green pentad are confounded on the angle $%
2\cdot \phi _3$ in their space under the Cartesian frame rotation about
Z-axis on this angle. Nevertheless, the triplet

\[
\left\{ 
\begin{array}{c}
\zeta ^x+\eta ^x+\theta ^x \\ 
-\zeta ^y+\eta ^y-\theta ^y \\ 
\zeta ^z+\eta ^z+\theta ^z
\end{array}
\right\} 
\]

behaves like the vector:

\[
\begin{array}{c}
\left( \zeta ^x+\eta ^x+\theta ^x\right) ^{\prime }= \\ 
=\left( \zeta ^x+\eta ^x+\theta ^x\right) \cdot \cos \left( 2\cdot \phi
_3\right) +\left( -\zeta ^y+\eta ^y-\theta ^y\right) \cdot \sin \left(
2\cdot \phi _3\right) , \\ 
\left( -\zeta ^y+\eta ^y-\theta ^y\right) ^{\prime }= \\ 
=\left( -\zeta ^y+\eta ^y-\theta ^y\right) \cdot \cos \left( 2\cdot \phi
_3\right) -\left( \zeta ^x+\eta ^x+\theta ^x\right) \cdot \sin \left( 2\cdot
\phi _3\right) \mbox{,} \\ 
\left( \zeta ^z+\eta ^z+\theta ^z\right) ^{\prime }=\zeta ^z+\eta ^z+\theta
^z\mbox{.}
\end{array}
\]

This triplet is denoted as the hadron monad ({Confinement}).

\section{OTHER ROTATIONS OF \protect\\THE CARTESIAN FRAME}

The Cartesian frame rotation about Y-axis:

Let $\phi _2$ be any real number and:

\[
U_y=-i\cdot \beta ^z\cdot \beta ^x\mbox{, }Q_y\left( \phi _2\right) =\cos
\left( \phi _2\right) \cdot E+i\cdot \sin \left( \phi _2\right) \cdot U_y%
\mbox{.} 
\]

Let for all $\vartheta $: $\vartheta ^{\prime }=Q_y\left( \phi _2\right)
^{\dagger }\cdot \vartheta \cdot Q_y\left( \phi _2\right) $.

In this case for light pentad:

\[
\begin{array}{c}
\beta ^{x\prime }=\beta ^x\cdot \cos \left( 2\cdot \phi _2\right) -\beta
^z\cdot \sin \left( 2\cdot \phi _2\right) \mbox{,} \\ 
\beta ^{y\prime }=\beta ^y \\ 
\beta ^{z\prime }=\beta ^z\cdot \cos \left( 2\cdot \phi _2\right) +\beta
^x\cdot \sin \left( 2\cdot \phi _2\right) \mbox{,} \\ 
\begin{array}{c}
\gamma ^{0\prime }=\gamma ^0\mbox{,} \\ 
\beta ^{4\prime }=\beta ^4\mbox{.}
\end{array}
\end{array}
\]

For chromatic pentads:

\[
\begin{array}{c}
\zeta ^{x\prime }=\zeta ^x\cdot \cos \left( 2\cdot \phi _2\right) -\theta
^z\cdot \sin \left( 2\cdot \phi _2\right) \mbox{,} \\ 
\theta ^{z\prime }=\theta ^z\cdot \cos \left( 2\cdot \phi _2\right) +\zeta
^x\cdot \sin \left( 2\cdot \phi _2\right) \mbox{,} \\ 
\theta ^{x\prime }=\theta ^x\cdot \cos \left( 2\cdot \phi _2\right) -\zeta
^z\cdot \sin \left( 2\cdot \phi _2\right) \mbox{,} \\ 
\zeta ^{z\prime }=\zeta ^z\cdot \cos \left( 2\cdot \phi _2\right) +\theta
^x\cdot \sin \left( 2\cdot \phi _2\right) \mbox{,} \\ 
\zeta ^{y\prime }=\zeta ^y\mbox{,} \\ 
\theta ^{y\prime }=\theta ^y\mbox{,} \\ 
\gamma _\zeta ^{0\prime }=\gamma _\zeta ^0\cdot \cos \left( 2\cdot \phi
_2\right) -\gamma _\theta ^0\cdot \sin \left( 2\cdot \phi _2\right) \mbox{,}
\\ 
\gamma _\theta ^{0\prime }=\gamma _\theta ^0\cdot \cos \left( 2\cdot \phi
_2\right) +\gamma _\zeta ^0\cdot \sin \left( 2\cdot \phi _2\right) \mbox{,}
\\ 
\zeta ^{4\prime }=\zeta ^4\cdot \cos \left( 2\cdot \phi _2\right) -\theta
^4\cdot \sin \left( 2\cdot \phi _2\right) \mbox{,} \\ 
\theta ^{4\prime }=\theta ^4\cdot \cos \left( 2\cdot \phi _2\right) +\zeta
^4\cdot \sin \left( 2\cdot \phi _2\right) \mbox{,} \\ 
\gamma _\eta ^{0\prime }=\gamma _\eta ^0\mbox{,} \\ 
\eta ^{4\prime }=\eta ^4\mbox{,} \\ 
\eta ^{y\prime }=\eta ^y\mbox{.}
\end{array}
\]

For the hadron monad:

\[
\begin{array}{c}
\left( \zeta ^x+\eta ^x+\theta ^x\right) ^{\prime }=\left( \zeta ^x+\eta
^x+\theta ^x\right) \cdot \cos \left( 2\cdot \phi _2\right) -\left( \zeta
^z+\eta ^z+\theta ^z\right) \cdot \sin \left( 2\cdot \phi _2\right) \mbox{,}
\\ 
\left( -\zeta ^y+\eta ^y-\theta ^y\right) ^{\prime }=\left( -\zeta ^y+\eta
^y-\theta ^y\right) \mbox{,} \\ 
\left( \zeta ^z+\eta ^z+\theta ^z\right) ^{\prime }=\left( \zeta ^z+\eta
^z+\theta ^z\right) \cdot \cos \left( 2\cdot \phi _2\right) +\left( \zeta
^x+\eta ^x+\theta ^x\right) \cdot \sin \left( 2\cdot \phi _2\right) \mbox{.}
\end{array}
\]

The Cartesian frame rotation about X-axis:

Let $\phi _1$ be any real number and:

\[
U_x=-i\cdot \beta ^y\cdot \beta ^z\mbox{, }Q_x\left( \phi _1\right) =\cos
\left( \phi _1\right) \cdot E+i\cdot \sin \left( \phi _1\right) \cdot U_x%
\mbox{.} 
\]

Let for all $\vartheta $: $\vartheta ^{\prime }=Q_y\left( \phi _1\right)
^{\dagger }\cdot \vartheta \cdot Q_y\left( \phi _1\right) $.

In this case for light pentad:

\[
\begin{array}{c}
\beta ^{x\prime }=\beta ^x \\ 
\beta ^{y\prime }=\beta ^y\cdot \cos \left( 2\cdot \phi _1\right) +\beta
^z\cdot \sin \left( 2\cdot \phi _1\right) \mbox{,} \\ 
\beta ^{z\prime }=\beta ^z\cdot \cos \left( 2\cdot \phi _1\right) -\beta
^y\cdot \sin \left( 2\cdot \phi _1\right) \mbox{,} \\ 
\begin{array}{c}
\gamma ^{0\prime }=\gamma ^0\mbox{,} \\ 
\beta ^{4\prime }=\beta ^4\mbox{.}
\end{array}
\end{array}
\]

For chromatic pentads ({QCD}):

\[
\begin{array}{c}
\eta ^{y\prime }=\eta ^y\cdot \cos \left( 2\cdot \phi _1\right) +\theta
^z\cdot \sin \left( 2\cdot \phi _1\right) \mbox{,} \\ 
\theta ^{z\prime }=\theta ^z\cdot \cos \left( 2\cdot \phi _1\right) -\eta
^y\cdot \sin \left( 2\cdot \phi _1\right) \mbox{,} \\ 
\theta ^{y\prime }=\theta ^y\cdot \cos \left( 2\cdot \phi _1\right) -\eta
^z\cdot \sin \left( 2\cdot \phi _1\right) \mbox{,} \\ 
\eta ^{z\prime }=\eta ^z\cdot \cos \left( 2\cdot \phi _1\right) +\theta
^y\cdot \sin \left( 2\cdot \phi _1\right) \mbox{,} \\ 
\eta ^{x\prime }=\eta ^x\mbox{,} \\ 
\theta ^{x\prime }=\theta ^x\mbox{,} \\ 
\gamma _\eta ^{0\prime }=\gamma _\eta ^0\cdot \cos \left( 2\cdot \phi
_1\right) +\gamma _\theta ^0\cdot \sin \left( 2\cdot \phi _1\right) \mbox{,}
\\ 
\gamma _\theta ^{0\prime }=\gamma _\theta ^0\cdot \cos \left( 2\cdot \phi
_1\right) -\gamma _\eta ^0\cdot \sin \left( 2\cdot \phi _1\right) \mbox{,}
\\ 
\eta ^{4\prime }=\eta ^4\cdot \cos \left( 2\cdot \phi _1\right) -\theta
^4\cdot \sin \left( 2\cdot \phi _1\right) \mbox{,} \\ 
\theta ^{4\prime }=\theta ^4\cdot \cos \left( 2\cdot \phi _1\right) +\eta
^4\cdot \sin \left( 2\cdot \phi _1\right) \mbox{,} \\ 
\gamma _\zeta ^{0\prime }=\gamma _\zeta ^0\mbox{,} \\ 
\zeta ^{4\prime }=\zeta ^4\mbox{,} \\ 
\zeta ^{y\prime }=\zeta ^y\mbox{.}
\end{array}
\]

For the hadron monad:

\[
\begin{array}{c}
\left( \zeta ^x+\eta ^x+\theta ^x\right) ^{\prime }=\left( \zeta ^x+\eta
^x+\theta ^x\right) \mbox{,} \\ 
\left( -\zeta ^y+\eta ^y-\theta ^y\right) ^{\prime }= \\ 
=\left( -\zeta ^y+\eta ^y-\theta ^y\right) \cdot \cos \left( 2\cdot \phi
_1\right) +\left( \zeta ^z+\eta ^z+\theta ^z\right) \cdot \sin \left( 2\cdot
\phi _1\right) \mbox{,} \\ 
\left( \zeta ^z+\eta ^z+\theta ^z\right) ^{\prime }= \\ 
=\left( \zeta ^z+\eta ^z+\theta ^z\right) \cdot \cos \left( 2\cdot \phi
_1\right) -\left( -\zeta ^y+\eta ^y-\theta ^y\right) \cdot \sin \left(
2\cdot \phi _1\right) \mbox{.}
\end{array}
\]

For all the Cartesian frame rotations - the chromatic pentads are
confounded, but the hadron monad behaves like the vector.

Therefore, the hadron current vector has got the following Cartesian
coordinates:

\[
\begin{array}{c}
j_x=\Psi ^{\dagger }\cdot \left( \zeta ^x+\eta ^x+\theta ^x\right) \cdot
\Psi \mbox{,} \\ 
j_y=\Psi ^{\dagger }\cdot \left( -\zeta ^y+\eta ^y-\theta ^y\right) \cdot
\Psi \mbox{,} \\ 
j_z=\Psi ^{\dagger }\cdot \left( \zeta ^z+\eta ^z+\theta ^z\right) \cdot
\Psi \mbox{.}
\end{array}
\]

\section{OTHER ROTATIONS}

Let us construct the rotation in the red and green pentads space like $U_x$, 
$U_y$, $U_z$:

Let $U_0=-i\cdot \gamma _\zeta ^0\cdot \gamma _\eta ^0$. Turned out to be,
that .$U_0=U_z$. Hence, this rotation is the Cartesian frame rotation. The
same for the other rotations into the chromatic pentads space.

But the rotation with the light and the chromatic pentads does not exist.

\section{RESUME}

Single light Clifford pentad and three chromatic ones define the particles
properties.

The weak isospin transformation corresponds to the rotation of the
antidiogonal Clifford matrices space.

The Newtonian gravity principle is the result of the sectioning of the
5-dimensional space on the 3-dimensional subspace $\Re $ and the
2-dimensional subspace $\grave A$.

The Cartesian frame rotations confound the chromatic pentads, but some these
pentads combination exists, which behaves as the vector under these
rotations.

\end{document}